\documentstyle[12pt]{article}
\def\qed{\nobreak\kern 1em \vrule height .5em width .5em depth 0em}
\def\vbar{\mathchoice{\vrule height6.3ptdepth-.5ptwidth.8pt\kern-.8pt}
   {\vrule height6.3ptdepth-.5ptwidth.8pt\kern-.8pt}
   {\vrule height4.1ptdepth-.35ptwidth.6pt\kern-.6pt}
   {\vrule height3.1ptdepth-.25ptwidth.5pt\kern-.5pt}}

\def\references
     {\vskip-\lastskip
    \vskip36pt plus12pt minus12pt
    \bigbreak
    \vfill\eject
     {\noindent \bf References\par}
      \parindent=0pt
      \bigskip}
\def\figures
   {\vskip-\lastskip
    \vskip36pt plus12pt minus12pt
    \bigbreak
    \noindent{\bf Figure Captions \par}
    \nobreak
    \bigskip
    \noindent}
\def\ack
   {\vskip-\lastskip
    \vskip36pt plus12pt minus12pt
    \bigbreak
    \noindent{\bf Acknowledgments\par}
    \nobreak
    \bigskip
    \noindent}
\def\date
   {\noindent Date: \today\par
    \medskip}
\def\refjl#1#2#3#4
   {\hangindent=16pt
    \hangafter=1
    \rm #1
   {\frenchspacing\sl #2
    \bf #3}
    #4\par}
\def\refbk#1#2#3
   {\hangindent=16pt
    \hangafter=1
    \rm #1
   {\frenchspacing\sl #2}
    #3\par}
\def\numrefjl#1#2#3#4#5
   {\parindent=40pt
    \hang
    \noindent
    \rm {\hbox to 30truept{\hss #1\quad}}#2
   {\frenchspacing\sl #3\/
    \bf #4},
    #5\par\parindent=16pt}
\def\numrefbk#1#2#3#4
   {\parindent=40pt
    \hang
    \noindent
    \rm {\hbox to 30truept{\hss #1\quad}}#2
   {\frenchspacing\sl #3\/}
    #4\par\parindent=16pt}
\newcount\subno      %number of subsection
\newcount\subsubno   %number of subsubsection
\newcount\appno      %appendix number
\newcount\appeqno    %appendix number
\newcount\tableno    %table number
\newcount\figureno   %figure number
\def\section#1
   {\vskip0pt plus.1\vsize\penalty-250
    \vskip0pt plus-.1\vsize\vskip24pt plus12pt minus6pt
    \subno=0 \subsubno=0
    \addtocounter{section}{1}
    \noindent {\bf \thesection. #1\par}
    \bigskip
    \noindent}
\def\subsection#1
   {\vskip-\lastskip
    \vskip24pt plus12pt minus6pt
    \bigbreak
    \global\advance\subno by 1
    \subsubno=0
    \noindent {\sl \thesection. \the\subno. #1\par}
    \nobreak
    \medskip
    \noindent}
\def\appendix#1
   {\vskip0pt plus.1\vsize\penalty-250
    \vskip0pt plus-.1\vsize\vskip24pt plus12pt minus6pt
    \subno=0
    \global\advance\appno by 1
    \noindent {\bf Appendix  #1\par}
    \bigskip
    \noindent}
\def\subappendix#1
   {\vskip-\lastskip
    \vskip36pt plus12pt minus12pt
    \bigbreak
    \global\advance\subno by 1
    \noindent {\sl \the\appno.\the\subno. #1\par}
    \nobreak
    \medskip
    \noindent}
\begin{document}
%
% Title Page
%
\def\footstrut{\baselineskip 12pt}
\hfuzz=5pt
\baselineskip 12pt plus 2pt minus 2pt
\vskip 24pt
\centerline{\bf WEAK DISORDER EXPANSION FOR THE ANDERSON MODEL ON A TREE}
\vskip 36pt
\centerline{J. Miller $^1$ \ and \  B. Derrida $^{1,\; 2}$  }
\vskip 12pt
\centerline{$^1$ \it
Service de Physique Th\'eorique, }
\centerline{\it CE Saclay F--91191 Gif-sur-Yvette Cedex, France}
\vskip 12pt
\centerline{ $^2$ \it   Laboratoire de Physique Statistique,
}
\centerline{\it Ecole Normale Sup\'erieure, 24 rue Lhomond,}
\centerline{\it 75231 Paris Cedex 05, France}
\vskip 36pt
\centerline{\bf ABSTRACT}
%\magnification=1200
\baselineskip 24pt plus 4pt minus 4pt

We show how certain properties of the Anderson model
on a tree are related to the solutions of a  non-linear integral
equation. Whether  the wave function is extended or localized, for example,
corresponds to whether or not  the equation has a complex solution.
We show how the equation can be solved in a weak disorder expansion.
We find that, for
small disorder strength $\lambda$, there is an energy $E_c(\lambda )$
above which the density of states and the conducting properties
vanish to all orders in perturbation theory. We compute
perturbatively the position of the line $E_c(\lambda )$
which begins, in the limit of
zero disorder, at the band edge of the pure system.
Inside the band of the pure system the density of states and
conducting properties can be computed perturbatively.
This expansion breaks down near $E_c(\lambda )$ because of small denominators.
We show how it can be resummed by choosing the appropriate scaling of the
energy. For energies greater than $E_c(\lambda )$ we show that non-perturbative
effects contribute to the density of states but have been unable tell
whether they also contribute to the conducting properties.

\vskip 0.5in
\centerline{To  be submitted to:
{\it  Journal of Statistical Physics } }
\vskip 0.25in
\date
\noindent Short Title: {\bf  Anderson Model on a Cayley tree
}\\
\medskip
\noindent PACS No: 71.30 71.50 J \\
\medskip
\medskip
\noindent  Key Words: localisation, tree, weak disorder expansion, mobility
edge
\medskip
\newpage
\baselineskip=18pt plus 3pt minus 2pt
\section{Introduction}
  Anderson localisation [1], or the study
of transport properties of
 a
 quantum particle in a random potential,  is one of
the most important problems in the theory of disordered systems [2,3].
In one and two dimensions an arbitrarily small
random potential suffices to localise all  energy eigenstates.
In three and higher dimensions both localised and
extended states can exist:
strong disorder or energies far from the band
center give rise to localised states  whereas weak disorder and
energies close to the band center   produce  extended states.
Extended and localised states are separated by a line in the
energy-strength of disorder plane, the mobility edge. The location of the
mobility edge
is a question of fundamental interest [4].

As usual in statistical mechanics, the simplest cases
one can consider are mean field models.  The most extensively studied
mean field model of
localisation is the Anderson model on a tree
 [5-12]. Various approaches have been developped, based in particular on self
 energy calculations [4,5] or on supersymmetry [8,9,13], which reduce
the problem to a non-linear integral equation [5-9].
This integral equation, however, is complicated and the position
of the mobility edge cannot be determined without
recourse to some kind of approximation.
Several works tried to overcome this difficulty by considering
simplified versions of the model on a tree [14,15].

In the present paper, we reconsider the Anderson problem on a tree. We first
give a  derivation of the integral equation
to be solved which, although completely equivalent to, is, we think,   more
intuitive than previous derivations.
The system insulates or conducts
depending on whether
 the integral equation possesses real or complex solutions.
 We  try to solve this equation in the limit of  weak disorder  using a method
[16]  which generalises previous weak disorder calculations in one dimension
[17].

One interesting outcome of this approach  is the existence of a line
$E_c(\lambda)$ in the
$E, \lambda$  plane ($E$ is the energy and $\lambda$ measures the strength of
disorder)
beyond which the integrated density of states and the conducting properties
vanish to
all orders in perturbation theory. This line tends to the band edge of the pure
system, $E= 2 \sqrt{K}$ (where $K+1$ is the coordination number of the tree) in
the limit of zero disorder. We can show that for energies greater than
$E_c(\lambda)$,  non-per
turbative contributions to the density of states  make it non-zero.
We have not, however, been able to
determine whether non-perturbative effects also contribute to the conducting
properties.  The question is of particular interest  because
Abou-Chacra and Thouless [5] predict that the mobility edge
 tends to $E = K+1$ rather than to the band edge, $E = 2 \sqrt{K}$
 in the limit of zero disorder.

The paper is organised as follows:
In section 2, we derive the non-linear integral equation satisfied by the
distribution
 $P(R)$ of a Riccatti variable $R$, defined  to be the ratio of
the wave function at adjacent sites on the lattice.
We show how the solution of this non-linear integral equation gives the
integrated density of states and why the existence
of a complex solution is related to the existence of extended states.
In section 3, we discuss the pure system, i.e. the problem in absence of
disorder. In section 4, we show how a weak disorder expansion
can be performed for energies inside the band of the pure system. We find that
in the presence
of  weak disorder   the system conducts.
In section 5, we extend the weak disorder expansion to the neighborhood of
the band edge. We obtain within this perturbative approach an expression
for the mobility edge $E_c(\lambda)$ in powers of the strength $\lambda$ of the
disorder. For $E >E_c(\lambda)$, the integrated density of states vansihes to
all orders in $\lambda$, although it is known that for distributions of the
potential with  unbou
nded support, it never vanishes  [18-20].
In section 6, we discuss the origin of non-perturbative effects
for energies outside the band of the pure system.
Lastly, in section 7, we describe a numerical method to obtain the
mobility edge, and we compare the results of this approach with
the prediction of section 5 and with an exactly soluble case where the
Ricatti variables are independent.

\section{ Formulation of the Problem}

We consider a tight binding model on a Cayley tree  of $N$ sites  (see figure
1) with
a random potential $V_i$ at each site $i$ of the lattice. The potentials $V_i$
are independent random variables  governed by a probability distribution
$\rho(V)$ which we choose to have  zero mean ($\langle V_i \rangle =0$).
The Schr\"{o}dinger equation  reads
\begin{equation}
\sum_{j=1}^{K+1} \ \psi_j \ = \ E \ \psi_i-\lambda\  V_i\  \psi_i \ .
\label{sch}
\end{equation}
Here $\psi_i$ is the value of the wavefunction at
site $i$,
$\lambda$ is a parameter that controls the
strength of the random potential,
$E$
is the energy of the particle, and the sum is over the
$K+1$ neighbors  of the site $i$.
It is useful to  rewrite (\ref{sch}) as a recursion relation [16].
Call
the central site  of the tree $i_0$ and define a
  Ricatti variable $R_i$ on a site $i$  by $R_i=\psi_j/\psi_i$,
where $j$ is the neighbor of site $i$  closer to
$i_0$  on the tree. Dividing
 (\ref{sch}) through by $\psi_i$
and regrouping terms gives (figure 1)
\begin{equation}
R_i =E-\lambda V_i -\sum_{j=1}^K {1\over R_j}
\label{rec}
\end{equation}
for all sites except  $i_0$.
 This recursion allows one to calculate the $R_i$ associated to all  sites of
the tree except for the  site $i_0$ (where it is not defined) and except for
the
sites adjacent to the boundary, where the $R_i$  depend on the boundary
conditions and,
 as we will see later, should be chosen differently depending on the properties
we want to study (density of states or conducting properties).

On account of the random potential in  (\ref{rec}), the $R_i$  are random
variables  governed by a
 probability distribution.
The recursion (\ref{rec}) completely determines the probability distribution
$P_m(R_i)$ of an $R_i$  located  $m$ steps from the boundary of the tree once
the probability distribution  $P_1(R_i)$ of the $R_i$ on sites adjacent to the
boundary have been specified.
In what follows we will always choose boundary conditions in such a way that
the
$R_i$ on sites adjacent to the boundary are identically distributed with
distribution
$P_1(R)$. The symmetry of the tree then ensures that all the $R_i$ an equal
number $m$ of steps from the boundary are also identically distributed with a
probability distribution $P_m(R_i)$ that depends on the number $m$ one has to
iterate (\ref{rec}).
The recursion (\ref{rec}) induces the following recursion on the distributions
$P_m(R)$
\begin{eqnarray}
P_{m+1}(R) =   \int \prod_{j=1}^{K} P_m(R_j) dR_j \int \rho(V) dV \ \delta
\left(R - E +\lambda V + \sum_{j=1}^{K} \frac{1}{R_j} \right) \ .
\label{P2}
\end{eqnarray}
We shall assume in what follows that, for all the boundary conditions  we
consider
 ($P_1(R)$ concentrated on the real axis when we calculate the density of
states or  $P_1(R)$ concentrated on the lower  half complex plane with all the
$R_i$
having a negative imaginary part when we study the conduction properties), the
recursion (\ref{P2}) converges to  a limiting distribution $P(R)$ which
satisfies
\begin{eqnarray}
P(R) =   \int \prod_{j=1}^{K} P(R_j) dR_j \int \rho(V) dV \ \delta \left(R - E
+\lambda V + \sum_{j=1}^{K} \frac{1}{R_j} \right) \ .
\label{Pf}
\end{eqnarray}
Up to a change of variables,
this integral equation is
equivalent to the integral equations obtained in [5-9]. A similar equation
also exists for diluted lattices
[21-23].
The  particular limiting distribution to which (\ref{P2}) converges  might
depend on the initial $P_1(R)$.
We will see below that
the localised and the extended regimes  correspond to one of the two following
 situations:

\begin{description}
\item {\bf In the localised region}: \\
There is only one fixed distribution, $P_{\rm real}(R)$, which solves
(\ref{Pf}). This distribution is concentrated on
the real axis and  is stable, i.e.  it is the limit of the sequence  $P_m(R)$
obtained through the recursion (\ref{P2})
for any initial distribution $P_1(R)$ (all  initial distributions  concentrated
on the real axis as well as those concentrated initially in the complex plane
converge to this distribution $P_{\rm real}(R)$, so that even if the $R$ have
initially  some imaginary part, they become real under the iteration of
(\ref{rec})).
\item {\bf In the extended region}: \\
There exist two different fixed distributions, $P_{\rm real}(R)$  and
$P_{\rm complex}(R)$, which solve (\ref{Pf}).
The real distribution  $P_{\rm real}(R)$,  concentrated on the real axis,
 is the limit of the sequence  $P_m(R)$
when the initial distribution $P_1(R)$
  is concentrated  on the real axis. This distribution $P_{\rm real}(R)$ is
however
unstable against imaginary perturbations: a small imaginary component in
the $R_i$ on the boundary will not vanish under iteration of (\ref{rec}).
Instead, if, as  for the scattering situation described below,
the initial distribution $P_1(R)$ is concentrated in the lower  half complex
plane (so that
the initial $R$ have all a negative imaginary part),
$P_m(R)$  converges to a different distribution $P_{\rm complex}(R)$
concentrated in
the lower half plane. (Note that there exists also a third distribution
in the upper half plane symmetric to $P_{\rm complex}(R)$  but we won't
consider  it because
with the boundary conditions we use
all the $R_i$ are always either real or complex with negative imaginary parts.)
\end{description}
The two fixed distributions $P_{\rm real}(R)$ and
 $P_{\rm complex}(R)$  are both solutions of the fixed point equation
(\ref{Pf})
and a great deal of  what follows  is devoted to the study of
 these fixed distributions.

For choices  of $E$,  $\lambda$ and $\rho(V)$ such that (\ref{Pf}) has  both a
real
and a complex  fixed distribution (the extended phase),
$P_{\rm real}(R)$  and
$P_{\rm complex}(R)$
 are not independent. In  the appendix we show that the real fixed distribution
$P_{\rm real}(R)$  solution of (\ref{Pf}) is given in terms of
 $P_{\rm complex}(R)$  by
\begin{eqnarray}
P_{\rm real}(R) && =
{1\over \pi}
\int_{-\infty}^{\infty} dr \int_{0}^{\infty}ds
{s\over (R-r)^2+s^2}P_{\rm complex}(r-is) \nonumber \\
&& =
{-1\over \pi}
{\rm Im} \left\{
\int_{-\infty}^{\infty} dr \int_{0}^{\infty}ds
{1\over R-r + i s}P_{\rm complex}(r-is) \right\} \ .
\label{pnrc}
\end{eqnarray}
\\ \\
We now  address the question of the choice of the initial distribution $P_1(R)$
 and discuss how the fixed distributions $P_{\rm real}(R)$ and $P_{\rm
complex}(R)$ are related to the density of states and to the conducting
properties.
\\ \\
{\bf The density of states:}
\\
Let us  first discuss how the density of states can be obtained
for the tree geometry [16]  (see figure 1). We  want to calculate the
eigenvalues  with the boundary condition that
the wavefunction vanishes on the boundary of the tree.
With this boundary condition, the
 Schr\"{o}dinger equation for  a site $i$ adjacent to the boundary reads
\begin{equation}
\psi_j = E \psi_i - \lambda V_i \psi_i
\end{equation}
where $j$ is the only neighbor of site $i$ on the tree.
Dividing through by $\psi_i$  then gives
\begin{equation}
R_i=E-\lambda V_i \  .
\label{rbcr}
\end{equation}
where $R_i= \psi_j/\psi_i$.
The initial $R$ are real, so   they remain real under  iteration of (\ref{rec})
 and the invariant
measure $P(R)$ is concentrated on the real axis.

As discussed in [16], the equations (\ref{rec}) and (\ref{rbcr})
solve the Schr\"{o}dinger equation everywhere on the tree except on the central
site
$i_0$. In terms of the $R_i$, the Schr\"{o}dinger equation for the central site
$i_0$ reads
\begin{equation}
E -\lambda V_{i_0}-\sum_{j=1}^{K+1}{1\over R_j} =0
\label{ri0}
\end{equation}
where the sum runs over the $K+1$ neighbors $j$ of site $i_0$.
  All the $R_i$ are functions of $E$ and the values $E_\alpha$ of the energy
which satisfy  (\ref{ri0}) are the eigenenergies.

Expression (\ref{ri0})  contains all the information on the density of states
but
is not  easy to use. If, however,
 one multiplies (\ref{ri0}) by
the product of all the $R_i$ in the lattice [16], it becomes a polynomial  in
$E$ of degree $N$ where $N$ is the number of lattice sites and
 the coefficient of  highest degree  is 1
 (for a tree of depth $n$, the number of sites $N= ( K^n +K^{n-1}-2)/(K-1)$).
Therefore, one can
write
\begin{equation}
\prod_{\alpha = 1}^{N} (E - E_{\alpha}) =
 \left(E -\lambda V_{i_0}-\sum_{j=1}^{K+1}{1\over R_j}\right) \prod_{i\neq i_0}
R_i \ .
\label{prod}
\end{equation}
Both sides of this equation are polynomials in $E$ with real coefficients and
real roots. To extract the density of states, one can   take the logarithm of
this equality, for any complex value of the
energy $E$,
with the convention that the branch cut runs along   the real axis from $-
\infty$ to  the largest eigenvalue $E_\alpha$. When the energy approaches the
real axis at a certain value $E$ from above,
the imaginary part of the left hand side is just $\pi$ times the number of
eigenenergies larger than $E$ whereas the imaginary part of the right hand side
is equal to $\pi$ times the number of negative $R_i$  ($+1$
when  the term $E- \lambda V_{i_0} - \sum_j \frac{1}{R_j}$  is negative).
This is because all the $R_i$ as well as the term
 $E- \lambda V_{i_0} - \sum_j \frac{1}{R_j}$   have positive imaginary parts
when the energy $E$ is in
the upper complex plane
(see  (\ref{rec})  and(\ref{rbcr})).
Therefore the number $\Omega_n(E)$ of eigenvalues greater than $E$
(the integrated density of states) is given by
\begin{equation}
\Omega_n(E) = \Theta( \lambda V_{i_0} + \sum_{j=1}^{K+1} \frac{1}{R_j}- E ) +
\sum_{i\neq i_0} \Theta( -R_i)
\label{x10}
\end{equation}
Since the number of negative $R_i$ is equal to the number of nodes
of the wave function, we see that the equality (\ref{x10}) between
 the integrated density of states and the number of nodes
of the wave function, well known  in one dimension, remains valid for tree
structures as was already discovered by Dhar and Ramaswamy [24] in a  similar
calculation of the
eigenmodes of Eden trees.

For a tree of depth $n$ (with $N= (K^{n+1}+K^{n} -2 )/(K-1)$ sites), the
average
over disorder of $\Omega_n(E)$ is given by
\begin{eqnarray}
\langle \Omega_n(E) \rangle = \int \rho(V_{i_0}) d V_{i_0}
\int_{-\infty}^{\infty} \cdots \int_{-\infty}^{\infty} \  \prod_{j=1}^{K+1} \
P_{n}(R_j) \  d R_j \  \Theta( \lambda V_{i_0} + \sum_{j=1}^{K+1}
\frac{1}{R_j}- E )
 \nonumber \\
+ \   (K + 1) \sum_{m=1}^{n}   K^{n-m} \int_{-\infty}^0 P_m(R) dR \ .
\label{x11}
\end{eqnarray}
This expression can be  simplified by using the recursion (\ref{P2})
\begin{equation}
\langle \Omega_n(E) \rangle =   \int_{-\infty}^{\infty} \  P_{n}(R) \  dR
\int_{-\infty}^{\infty} \   P_{n+1}(R') \   dR' \ \Theta(  \frac{1}{R}-  R' )
 \nonumber \\
+ (K+1) \sum_{m=1}^{n}   K^{n-m} \int_{-\infty}^0 P_m(R) dR \ .
\label{Omega}
\end{equation}

The tree geometry has the pathology
that the
number of sites near the boundary
is proportional to the total number of sites
in the tree. Surface effects are therefore strong.  One can see this in
(\ref{Omega})
where   $P_1(R)$ is multiplied by the largest power of $K$.
To eliminate these boundary effects
and obtain an expression
for the behavior of the bulk,
 one can use a subtraction procedure.
Letting $F_n$ be an extensive quantity in a tree of
depth $n$, the quantity
$(F_n-KF_{n-1})/2$ is the value of $F$ per site far from
the boundary when $n$ is large [16]. To see this,
note that the
number of sites $m$ steps from the boundary
in a tree of depth $n$ equals $K$ times the number of
sites $m$ steps from the boundary in a tree of depth $n-1$.
The contributions to $F$ from sites $m$ steps from the boundary
are thus cancelled in the subtraction.
Under this subtraction, the number of sites we are left  with is
$(K^{n+1}+K^{n}-2)/(K-1) - K (K^{n}+K^{n-1}-2)/(K-1) = 2$ sites which are far
from the boundary.
  Thus dividing   the difference $F_n - K F_{n-1}$  by two
gives the value of $F$ per site  far from the boundary.
  Applying this subtraction to (\ref{Omega}) gives  the average integrated
density of states $\langle \omega(E)\rangle$ per site far from the boundary  in
the limit $n \rightarrow \infty$
\begin{eqnarray}
\langle \omega(E) \rangle =   -  \  \frac{ K-1}{2}\int_{-\infty}^{\infty} \
P_{\rm real}(R) \  dR  \int_{-\infty}^{\infty} \   P_{\rm real}(R') \   dR' \
\Theta(  \frac{1}{R}-  R' )
\nonumber \\
+ \  \frac{K+1}{2} \int_{-\infty}^0~ P_{\rm real}(R)  \ dR \ .
\label{omegar}
\end{eqnarray}
For values of $E$ and $\lambda$ such that there exists a complex fixed
distribution
$P_{\rm complex}(R)$, one can use the relation (\ref{pnrc}) between $P_{\rm
real}$ and
$P_{\rm complex}$ to express
$\langle \omega(E) \rangle$ in terms of this complex distribution (see the
appendix)
\begin{eqnarray}
\langle \omega(E) \rangle =   \frac{1}{\pi} {\rm Im}\left\{     \frac{
K-1}{2}\int \  P_{\rm complex}(R) \  dR  \int \   P_{\rm complex}(R') \   dR' \
\log( R'-  \frac{1}{R} )  \right.
\nonumber \\
 \left.- \  \frac{K+1}{2} \int P_{\rm complex}(R)  \ dR  \  \log(R) \right\}
\label{omegac}
\end{eqnarray}
\\
{\bf The conducting properties:}\\
We turn now to the relation
between $P(R)$ and the conducting properties of the system.
Imagine the  situation
shown in figure 2.
Attach a wire to each boundary
site of a branch of a tree as shown in figure 2. Suppose then that
one sends a plane wave in at the left and allows it to scatter
off the tree.
For a branch of finite depth, some of the
incoming wave will be reflected
and some will propagate through the branch to the wires
on the right.
Of interest is what happens when the depth of the branch
becomes large.
There are two possibilities: either   the
wave is entirely reflected or some of the wave propagates through
the tree into the wires on the right. In the former case,
 either there is
a gap in the energy spectrum or the
wavefunctions of the tree are localised, in the latter case
the wavefunctions are extended.
Now it is known that for
a potential with  unbounded support  (e.g.
a Gaussian)  [18-20] there are states at all energies.
In this case, a  reflection  amplitude of modulus 1
(complete reflection of the wave) implies that
the states at that energy are localised.
Therefore,
to determine whether at a given value
of $E$ and $\lambda$ the particle is localised,
it suffices to compute the reflection amplitude in this
experiment.

To relate the reflection amplitude to the Ricatti variables,
one needs to analyse what  happens at the boundary.
First
consider the Schr\"{o}dinger equation  for a site $a$ adjacent to the right
boundary
\begin{equation}
 \psi_b+\psi_c\label{ pwr} =
E \psi_a  - \lambda  V_a \psi_a
\end{equation}
 where $b$ and $c$ are the neighbors of $a$ on the tree
and in the wire respectively (figure 2).
In the wire the wavefunction is a plane wave
$\psi =e^{i k x}$ going to the right
(corresponding to no incoming flux from right infinity)
so that $\psi_c/\psi_a = e^{ik}$.
The Ricatti variable $R_a = \psi_b/\psi_a$ thus equals
\begin{equation}
R_a = E - \lambda V_a -e^{ik }.
\label{rbcc}
\end{equation}
The wave vector $k$ can be adjusted by putting
a uniform potential on the wire.
We note that the $R_a$ for  all  sites adjacent to the boundary have a negative
imaginary part and it is easy to check that the iteration (\ref{rec}) preserves
this property.

Starting with the boundary $R_a$ given by (\ref{rbcc}), we now iterate  the
recursion  (\ref{rec})
up to the first $R$ in the wire on the left side of the
tree (figure 2). At this boundary we have both an
incoming and an outgoing wave, so that the wave
function on the left wire has the form $\psi =e^{i k x}+r e^{-i k x}$
where $r$ is the reflection amplitude.
The reflection amplitude $r$ is therefore related to the Ricatti variable
$R_{d}$ by
\begin{equation}
R_{d}={\psi_{e}\over \psi_{d}}={e^{-i k}+re^{ik}\over {1+r}}\label{Rr}
\end{equation}
or  equivalently,
\begin{equation}
r=-\frac{R_d-e^{-i k}}{ R_d-e^{i k}}.\label{ rR}
\label{reflection}
\end{equation}
It is clear that if $R_d$ is real,
the numerator and the denominator
 are complex conjugates and so $|r|=1$. It is also easy to see that
if $R_d$ has a negative imaginary part,
$|r| < 1 $.
Thus
 if the imaginary parts
of the initial $R_i$ iterate to zero, the particle
is totally reflected  and if they do not, the particle  has
a non-zero transmission coefficient.
 To find  the mobility edge  it therefore
suffices to study the complex fixed distribution $P_{\rm complex}(R)$ and
to determine at which values of $E$ and $\lambda$
the probability of finding a non-zero imaginary part
of $R$ first vanishes.  \\ \\
 Remark: An idealized scattering situation like the one shown in
figure 2
can be used in other cases including
finite dimensional lattices
to decide whether a system is conducting or insulating.
 One could have for example an incoming wave on a wire attached to an internal
site and outgoing waves at the boundary. A   straightforward calculation  but
which would  require    new notations (that we will not present here)  would
show   that a reflec
tion amplitude smaller  than 1
in the scattering situation is  equivalent to the Green's function having a
non-vanishing
imaginary part when the energy $E$ tends to the real axis [25,26 and references
therein].

For arbitrary $\lambda$, $E$ and $\rho(V)$,
one does not know how to find the fixed distributions   that solve
 (\ref{Pf}) and in the following sections we will
show how certain quantities can be expanded in the limit
of weak disorder ($\lambda $ small). An exception is when the distribution
of the potential is Cauchy [5]
\begin{equation}
\rho(V) = \frac{1}{\pi} \ \frac{1}{V^2 + 1} \ .
\end{equation}
Using the fact that  sums of Cauchy random variables are also
Cauchy distributed and that the inverse of a Cauchy variable is Cauchy, it is
easy to
obtain the exact form of  $P_{\rm real}(R)$:
\begin{equation}
P_{\rm real}(R) = \frac{1}{\pi} \  \frac{b}{(R-a)^2 + b^2}
\label{PCauchy}
\end{equation}
where the parameters $a$ and $b$ are the values of the attractive fixed point
of
the following two dimensional map
\begin{eqnarray}
a_{n+1} = E - K \frac{a_n}{a_n^2 +b_n^2} \ \ \ \ \ ; \ \ \ \
b_{n+1} =  \lambda + K \frac{b_n}{a_n^2 + b_n^2}  \ .
\nonumber
\end{eqnarray}
Using (\ref{omegar}) one then finds  the following closed expression for
the integrated density of states $\langle \omega(E) \rangle$
\begin{equation}
\langle \omega(E) \rangle = \frac{K+1}{2 \pi}  \tan^{-1}\left( \frac{b}{a}
\right) \
- \  \frac{K-1}{2 \pi} \tan^{-1}\left( \frac {b}{a} \ \
\frac{a^2+b^2+1}{a^2+b^2-1}  \right) \ .
\end{equation}
Unfortunately, no one has to date been able to obtain  an exact expression
for the complex fixed distribution, even in the case of a Cauchy distributed
potential.
So  conduction properties such as the location of the mobility edge are not
known exactly  even for Cauchy disorder.
\section{The pure system}
Before proceeding to the  weak disorder expansions,
let us consider the case  of no disorder ($\lambda =0$). For zero disorder and
for our choices
of boundary conditions,
all the $R_i$ for the boundary sites are equal, and so
 the initial distribution $P_1(R)$ is a $\delta$ function.
In the absence of disorder, all the $P_k(R)$
computed from $P_1(R)$  through the recursion (\ref{P2})
are also $\delta$ functions concentrated at some value $A_k$
\begin{equation}
P_k(R) = \delta( R - A_k)
\label{delta}
\end{equation}
where the $A_k$ satisfy the following recursion
\begin{equation}
A_{k+1} = E - \frac{K}{A_k} \ .
\label{An}
\end{equation}
This recursion for $A_k$ has two fixed points which are real when $|E| >
2\sqrt{K}$
and complex conjugate when $ -2 \sqrt{K} < E < 2 \sqrt{K}$.
\\ \\
First, if
 \begin{equation}
 E > 2 \sqrt{K}
\label{Eout}
\end{equation}
the sequence $A_k$ given by the recursion (\ref{An})  always converges  to
the real fixed point $A  $ given by
\begin{equation}
A = \frac{E + \sqrt{E^2 -4K}}{2} \ .
\label{Areal}
\end{equation}
This implies that in the scattering situation described in section 2, the
initial complex $R_i$
become real under  iteration, and  hence the wave is completely reflected.
Furthermore,the integrated density of states is zero for this range of energy.
To see this, note that with the boundary condition (\ref{rbcr}) (i.e. $R_i =
E$), appropriate for the calculation of the density of states,
 $A_1 > A_2 > \cdots > A_n \cdots > A$, and so   all the $R$
are positive. Moreover $E - (K+1) / A_n $ is also  positive.  It then follows
from
expressions  (\ref{x10}) or (\ref{omegar}) that the integrated
density of states is zero for the range of energy (\ref{Eout}), meaning that
even for a finite tree,
there is no eigenenergy greater than $ 2 \sqrt{K}$ and
 so the range of energy (\ref{Eout}) is outside the band.
The case $E < - 2 \sqrt{K}$ is obviously symmetric.
\\ \\
 On the other hand, if
\begin{equation}
 -2 \sqrt{K} < E < 2 \sqrt{K}
\label{Ein}
\end{equation}
there exists a complex  fixed point of (\ref{An})
\begin{equation}
A   = \frac{E - i \sqrt{4 K - E^2}}{2} \ .
\label{Astar}
\end{equation}
meaning that there is a $\delta$-distribution concentrated at this point that
solves
 ({\ref{P2}).

If one starts with a real $A_1$  as in ({\ref{rbcr}) to compute the density of
states,
the sequence $A_k$ does not converge.  Letting the energy
$E= 2 \sqrt{K} \cos \theta$,  one finds
\begin{equation}
A_{n} = \sqrt{K} \  \frac{  \sin(n \theta + \theta)}{\sin (n \theta)} \ .
\end{equation}
The integrated density of states  for a tree of depth $n$  then follows
from  (\ref{x11}):
\begin{equation}
\Omega_n(E) = \Theta\left( \frac{ \sin n \theta - K \sin(n \theta + 2
\theta)}{\sin(n \theta + \theta)}  \right)
+ (K+1) \sum_{m=1}^n K^{n-m}  \ \Theta \left( \frac{  -  \sin(m \theta +
\theta)}{\sin(m \theta )}  \right) \ .
\end{equation}
The spectrum consists of a finite number of eigenvalues (as it should for a
finite system) with
huge degeneracies which reflect the symmetries of the tree.

If one starts with  a complex $A_1$, the sequence (\ref{An}) does not converge
either. It is
however easy to show that
\begin{equation}
\frac{ A_k - \sqrt{K} e^{i \theta}}{A_k -\sqrt{K} e^{-i \theta}} =
e^{-2(k-1) i \theta}
\frac{ A_1 - \sqrt{K} e^{i \theta}}{A_1 -\sqrt{K} e^{-i \theta}}.
\end{equation}
 From this explicit expression, we see that if $A_1$ is complex, $A_k$
remains  complex, and so the system is conducting. \\ \\
Remark: tt is a property particular to the pure system  that when  the $R_i$ at
the boundary are all equal, the sequence $A_n$ does not converge, and thus the
sequence $P_k(R)$ has no limit. As soon as one introduces disorder ($\lambda
\neq 0$), there are limiting distributions $P_{\rm real}(R)$ and $P_{\rm
complex}(R)$, which satisfy the fixed point (\ref{Pf}). These distributions, in
the limit $\lambda \rightarrow 0$,   also satisfy the fixed point equation
(\ref{Pf}). It is possible to find these distributions. The complex
distribution is a $\delta$-function concentrated at $A$ given by (\ref{Astar})
and the real distribution  is given by the relation (\ref{pnrc})  between
$P_{\rm real}(R)$ and
$P_{\rm complex}(R)$
\begin{equation}
P_{\rm real}(R) = \frac{1}{2 \pi} \  \frac{\sqrt{4 K - E^2}}{ R^2 - E R + K} \
{}.
\label{cauc}
\end{equation}
 It is interesting to notice that this fixed distribution (\ref{cauc}) is the
invariant measure of the map (\ref{An}). Using this fixed distribution, one
obtains the following expression for
the integrated density of states $\omega(E)$ per site far from the boundary
\begin{equation}
\omega(E) = \frac{K+1}{2 \pi} \theta - \frac{K-1}{2 \pi} \tan^{-1} \left(
\frac{K+1}{K-1} \tan \theta \right) \ .
\end{equation}

\section{Weak disorder expansion inside the band}
We turn now to  the  weak disorder expansions.
As the properties of the pure system are qualitatively different in the band
and
outside the band,  the weak disorder expansion requires rather different
techniques in these two cases. We
will see that  the neighborhood of the band edge must also be treated
separately.

In this section, we explain the small $\lambda$ expansion  for energies $E$
inside the
band of the pure system ($- 2 \sqrt{K} < E < 2 \sqrt{K}$).
In this energy range,  we saw that the  complex fixed distribution
solution of (\ref{P2})  is
a $\delta$-function
concentrated at the complex number  $A$
\begin{equation}
A  = \frac{ E - i \sqrt{4 K - E^2}}{2} \ .
\end{equation}
To  obtain the weak disorder expansion  we
 assume that for $\lambda$ small, the variables
$R$ which appear in the recursion
(\ref{rec}) have small fluctuations around $A$ (see figure 3
and remark 1 at the end of this section),
so that the distribution $P_{\rm complex}(R)$
remains concentrated around
$A$.
In accordance with this assumption [16], it is
convenient to make the following change of variables
\begin{equation}
R_i =  {A\over 1+\lambda B_i +\lambda^2 C_i + \cdots}   \ .
\label{expand}
\end{equation}
Here $B_i,C_i,\cdots$ are fluctuating quantities, the distributions of which
are assumed
to independent of $\lambda$.
Inserting (\ref{expand}) into (\ref{rec}) and  equating terms
 order by order in $\lambda$,
one finds that  the fluctuating parts  $B_i, C_i,\cdots$ of $R_i$ satisfy the
following recursions
\begin{eqnarray}
A B_i = V  + \frac{1}{A} \sum_{j=1}^K B_j
\ \ \ \ \  ; \ \ \ \ \
A C_i - A B_i^2 =  \frac{1}{A} \sum_{j=1}^K C_j
\end{eqnarray}
 and so on. Using these equations, it is  easy to compute the moments of the
fluctuating variables $B_i, C_i, \cdots$ in terms of $K$, $E$ and
the moments $\langle V^p\rangle$ of the random potential $V$ which we assume to
be finite.  For example
\begin{eqnarray}
\langle B \rangle = 0  \ \ \ \ \ \ ; \ \ \ \ \ \
\langle B^2 \rangle =\frac{ A^2 \langle V^2 \rangle}{A^4 - K } \ \ \ \ \ \ ; \
\ \ \
\langle C \rangle =\frac{ A^4 \langle V^2 \rangle}{(A^4 - K )(A^2 - K )}  \ .
\end{eqnarray}
Using this  weak disorder expansion one can   then calculate perturbatively
any quantity that can be expressed as a function of the $R_i$. For example the
real fixed point can be obtained from
(\ref{pnrc})
\begin{equation}
P_{\rm real}(R) =  \frac{-1}{\pi}{\rm Im} \left\{
\frac{1}{R-A} + \lambda^2 \langle V^2 \rangle \frac{A^3 (A^3 - K R)}{(R-A)^3
(A^2 - K )(A^4 - K )}  + O( \lambda^3) \right\} \ .
\end{equation}
The integrated density of states can also be calculated  using (\ref{omegac})
which relates $\omega(E)$  to the complex fixed point.
Substituting the perturbative expansion (\ref{expand}) of $R$
around the complex fixed point into (\ref{omegac}), expanding
in $\lambda$ and calculating the
necessary moments  we
find to order $\lambda^4$
\begin{eqnarray}
\omega(E)={-1\over \pi}{\rm Im}\left\{K \log A -{(K-1)\over 2}\log
(A^2-1)-\lambda^2 \langle V^2 \rangle {A^2\over 2(A^2-1)^2}-\lambda^3 \langle
V^3\rangle {A^3
\over 3(A^2-1)^3} \right.
\nonumber \\
 \left. -\lambda^4 \langle V^4 \rangle {A^4 \over 4(A^2-1)^4}+
\lambda^4 \langle V^2 \rangle^2 {A^4(3K-A^2-2A^4)(K+1)\over
4(A^2-K)(A^4-K)(A^2-1)^4}+
O(\lambda^5) \right\}
\label{omepert}
\end{eqnarray}
Higher order terms in $\lambda$ are  straightforward to calculate
but in practice the algebra required
to work out the necessary moments and correlations
becomes rather involved.

Conduction properties
can be calculated  in a similar way.
A question of basic interest  is whether
the wave function is localised or extended.
 From the discussion of section  2
we know that  the wavefunction is  extended  if there exists
a  complex fixed distribution concentrated in the lower half plane.
One way of determining whether such a complex fixed distribution exists is
to calculate the magnitude of the reflection amplitude and see whether it is
less than
(extended) or equal to (localised) 1.
The existence of the complex fixed distribution can be determined more easily
however by considering $\langle {\rm Im} R \rangle$. Since the complex fixed
distribution
is concentrated in the lower half plane, a necessary and sufficient condition
for it to exist
is  for
 $\langle {\rm Im} R \rangle$  to be non-zero.
Perturbatively we have
\begin{eqnarray}
 &&\langle {\rm Im} R \rangle =\langle {\rm Im}\left\{A(1 -\lambda B-\lambda^2
C+ \lambda^2 B^2 ...\right\} \rangle
\label{imaginary}
 \\
 && = {\rm Im} \left\{
A -\lambda^2 \langle V^2\rangle
{K A^3 \over ( A^2-K)(A^4-K)}
+ \lambda^3 \langle V^3 \rangle \frac{ K A^4 (A^4 +K)}{(A^2-K)(A^4-K)(A^6-K)}
+O(\lambda^4)
\right\}
\nonumber
\end{eqnarray}
\\ \\
Remark 1: Equation (\ref{imaginary}) indicates that for energies $E$ inside the
band of the pure system,
 weak disorder leaves the eigenfunctions extended. This result, obtained for
arbitrary $K$ including
   $K=1$  where the tree becomes one dimensional, apparently
contradicts
 the  well known fact that  weak disorder
localises all the eigenfunctions  in one dimension [27].
The reason  the above calculation does not apply when $K=1$ is that the
assumption that  the iteration (\ref{P2}) converges for weak disorder to a
fixed complex distribution close to
the fixed point $A  $ is not valid. This can be easily seen in the  weak
disorder expansion.
Assume that some $R_i$ obtained after $n$ iterations has the form $R_i =A  /(1+
\lambda B_i^{(n)}+ \cdots)$.
Clearly one has
\begin{equation}
B_i^{(n)} = \frac{V_i}{A  } + \frac{1}{A^{2}} \sum_{j=1}^{K} B_j^{(n-1)} \ .
\end{equation}
This implies the following recursion for $\langle | B^{(n)}|^2 \rangle$
\begin{equation}
\langle | B^{(n)} |^2 \rangle = \frac{\langle V^2 \rangle}{K}
+ \frac{\langle |B^{(n-1)}|^2 \rangle}{K}
\label{x42}
\end{equation}
where we have used the equality $|A  | = \sqrt{K}$.
We see from (\ref{x42}) that  for $K>1$ the fluctuations $\langle | B^{(n)} |^2
\rangle $ saturate as $n$ increases, meaning that  an initial distribution
concentrated in the neighborhood of $A  $ remains concentrated near $A  $. On
the other hand,
for $K=1$, the fluctuations grow  linearly with $n$ so that  the recursion does
not converge to a fixed
distribution
close to $A  $.
\\ \\
Remark 2:  The case $K = 1$ (one dimension) is  special in other respects.  In
the band, at energy $E = 2 \sqrt{K} \cos \theta$, the fixed point $A$ is $A=
\sqrt{K} e^{-i \theta}$.  Hence
 for $K=1$, there are a large number of values of $\theta$ at which the weak
disorder  expansion (\ref{omepert})  contains small denominators [17,28-31].
This is not  the case for $K > 1$ where only the band edges ($\theta=0$ and
$\theta = \pi$) seem to give rise to small denominators.

\section{Near the Band Edge}

As $E$ approaches  the band edge
of the pure system  $2 \sqrt{K}$ from below at fixed  $\lambda$,
terms like $(A^2 -K)^{-1}$
diverge, and the above perturbation theory
breaks down.
Perturbative information about the complex fixed point can, however,
still be obtained by choosing the appropriate scaling
of
the energy with $\lambda$.
Let
\begin{equation}
E=\sqrt K(2 -a \lambda^2)
\label{Ea}
\end{equation}
where $a$ is of order one.    For these energies,
 $A$ is given  by $\sqrt{K} (1 -i \lambda \sqrt{a} + O(\lambda^2)) $.
Substituing
this in the expansions
(\ref{omepert}) and (\ref{imaginary}) one finds
\begin{equation}
\langle \omega(E) \rangle =  \lambda^3 \frac{2 K(K+1)}{3 \pi (K-1)^2} \left(
a^{3/2} + \frac{3\langle V^2 \rangle a^{1/2}}{2(K-1)}
+ \frac{ 3 \langle V^2 \rangle^2 }{8(K-1)^2 a^{1/2}} +\cdots \right)
\end{equation}
 and
\begin{equation}
\langle {\rm Im } R \rangle =- \lambda \sqrt{K} \left( a^{1/2} +\frac{\langle
V^2 \rangle  }{2(K-1) a^{1/2}} +\cdots \right) \ .
\end{equation}
 From this we see that the leading term in a small $\lambda$ expansion with $a$
fixed as $\lambda \rightarrow 0$ corresponds to resumming an infinite series in
the original expansions (\ref{omepert}) and (\ref{imaginary}).

For the range of energies (\ref{Ea}), let us make  the following change of
variable
\begin{equation}
R_i={\sqrt K\over 1-\phi_i} \ .
\end{equation}
Substituing this expression in (\ref{rec}), we see that  the $\phi_i$ are of
order $\lambda$. Expanding both sides of (\ref{rec})  yields
\begin{equation}
\phi_i+\phi_i^2+\phi_i^3+...=-{\lambda V\over \sqrt
K}-a\lambda^2+\frac{1}{K}\sum_{j=1}^K\phi_j
\label{phi}
\end{equation}
The moments of $\phi$
can then be calculated by taking successive integer powers of  this equation.
To lowest order one obtains
\begin{equation}
\langle \phi^2 \rangle = - a\lambda^2 \ \ \ ;\ \ \
\langle \phi \rangle^2 = - \lambda^2 \left( a + \frac{\langle V^2 \rangle}{K-1}
\right) \ \ \ ; \ \ \  \langle \phi \rangle^3 = O( \lambda^3)
\label{phiband}
\end{equation}
and so forth.
Just as in the last section we can now
calculate the average of functions with respect to the complex
fixed point perturbatively.
For example,
one finds using (\ref{omegac})
for  the integrated density of states
\begin{equation}
\langle \omega(E) \rangle=\lambda^3 \frac{2 K (K+1)}{3 \pi (K-1)^2 }\left(a +
\frac{\langle V^2 \rangle}{K-1}\right)^{3/2}+ O(\lambda^4) \ \ \ \ {\rm for} \
\ \ a > - \frac{\langle V^2 \rangle}{K-1}
\label{omband}
\end{equation}
and
$
\langle \omega(E) \rangle= 0
$  to order $\lambda^3$ for
$ a < - \langle V^2 \rangle / (K-1) $.
Both (\ref{phiband}) and (\ref{omband}) agree with(\ref{omepert}) and
({\ref{imaginary}) when $a$ becomes large as it should and (\ref{phiband})
and(\ref{omband}) are expected
to represent the ressumed series of the most singular terms in the expansions
(\ref{omepert}) and (\ref{imaginary}).

 One might wonder how this result is modified by higher order terms.
If one decides to calculate the integrated density of states or any other
quantity
to a given order  $p$ in $\lambda$, one needs to calculate  the first $p$
moments
  $\phi$  to order $\lambda^p$. To do so, one takes the successive powers of
(\ref{phi}) and
one averages. This gives relationships between the first $p$ moments of $\phi$.
If one uses these relations to express all the moments in terms of the first
moment
$\langle \phi \rangle$, one ends up with a polynomial equation in $\phi$
of degree $p$ which has real coefficients depending on $\lambda$, $a$ and the
moments of the potential $V$. As one varies $a$ (i.e. the energy $E$), one
finds that
two complex roots of this polynomial become real at a certain critical value
$E_c$
which  to fourth order  in $\lambda$ reads:
\begin{equation}
\frac{E_c}{\sqrt{K}} = 2  +
 \lambda^2  \frac{ \langle V^2 \rangle }{K-1}
+ \lambda^3  \frac{\sqrt{K} \langle V^3 \rangle }{(K-1)^2}
+ \lambda^4  \frac{K \langle V^4 \rangle }{(K-1)^3}
- \lambda^4  \frac{( 9 K^2 -14 K -3) \langle V^2 \rangle^2 }{4(K-1)^4}
+ O( \lambda^5) \ .
\label{Ec}
\end{equation}
If one pushes  this expansion to higher order in $\lambda$, one would find
higher order corrections to $E_c$ but  for $E > E_c$, $\langle \phi \rangle$
as well as all the higher moments of $\phi$ would be real, and this implies
through (\ref{omegac}) tha
t $\omega(E) = 0$ for $E> E_c$ to any order in $\lambda$.

This result  seems to contradict the well known fact that
the density of states never vanishes
for distributions of the random potential with unbounded support
(such as the Gaussian distribution) [18-20].
Our result can be reconciled with a non-zero density of states
at all energies $E$ only if
 there are small non-perturbative contributions
to the density of states
above $E_c$  which vanish to any order in $\lambda$. We  discuss this point  in
the next section.

If perturbation theory could be trusted,
it would follow from the discussion of section two that
the particle is localised
for energies above $E_c$
since $\langle \phi \rangle$ is real.
The perturbative calculation
therefore predicts  a mobility edge  at $E_c$   given by (\ref{Ec}).
This result could be modified by non-perturbative effects; unfortunately,
unlike the case for the density of states, we have not been able to
discover what these effects might be or how to calculate them.

\section{ Beyond the Band Edge}
In this section we consider energies larger than
$2\sqrt K$ and outside the scaling regime discussed
in the last section.
In this range of energy,  the $R_i$ converge to the real fixed point $A  $
given by (\ref{Areal}) when $\lambda =0 $.  For small $\lambda$, one expects
the $R_i$ to fluctuate
about  $A  $.   Expression (\ref{omegar}) for the density of states requires
however
 the knowledge of $P_{\rm real}(R)$ for values of $R$  far from $A  $. This is
a large deviation problem
since untypical values of $R$ are produced by untypical values of
the potential.    We are going to show  that
the real fixed distribution $P_{\rm real}(R)$ can
be calculated by a saddle point  method. We have also tried to calculate
 the complex
fixed   distribution in this range of energies by a similar approach
 but  failed and we are unable to tell whether the fixed
 complex distribution simply does not exist
or whether we have just not been able to find it.

Since the regions of $R$ which contribute to the density
of states corresponds to untypical values of $R$,
the shape of $P_{\rm real}(R) $ in these regions
  depends strongly on
the shape of $\rho(V)$. In what follows, we will  take  $\rho(V)$ to be
Gaussian
\begin{equation}
\rho (V)= \frac{1}{ \sqrt{2 \pi}} \exp \left( - \ \frac{ V^2}{2} \right) .
\end{equation}
The equation to be solved for $P_{\rm real}(R) $ is is then
\begin{equation}
P(R)=\int \frac{d V}{ \sqrt{2 \pi}} \exp \left({-V^2\over 2}\right)
\prod_{i=1}^K \  dR_i  \ P(R_i)
\ \delta \left(R-E+\lambda V+\sum_{i=1}^K{1\over R_i}\right)
\label{P3}
\end{equation}
Now suppose  that $\lambda$  is small and that $R$ has fluctuations of order
$\lambda$
around $A  $. If we write $R=A  +\lambda \phi$, we get from (\ref{rec})
\begin{equation}
\phi =-\lambda V+{1\over A^{2}}\sum_i\phi_i \ .
\end{equation}
Since $V$ is Gaussian distributed, it follows that
$\phi$ is also Gaussian with zero mean and
$\langle\phi^2\rangle=\langle V^2 \rangle / (1- K/ A^{4})$.
In terms of the variable $R$,
\begin{equation}
P(R) \simeq  \frac{1}{\sqrt{2 \pi \lambda^2 \langle \phi^2 \rangle}}
\exp{-{(R-A)^2\over 2  \lambda^2 \langle \phi^2 \rangle}}
\end{equation}
valid for $R-A  $ of order $\lambda$.
Given this expression  for $P(R)$ valid for small fluctuations
away from  $A  $, it is natural to
look for   the solution of (\ref{P3}) of the form
\begin{equation}
P(R)=Q(R)  \ \exp \left({F(R)\over \lambda^2} \right)
\label{PQF}
\end{equation}
where $F(R)$ is independent of $\lambda$.
Plugging this into (\ref{P3}) and integrating over $V$ gives a
saddle point equation for $F(R)$
\begin{equation}
F(R) = \max_{R_1,R_2, \cdots R_K} \left\{   - \ \frac{1}{2} \left( R - E +
\sum_{i=1}^K \frac{1}{R_i} \right)^2 + \sum_{i=1}^K F(R_i) \right\} \ .
\label{saddle}
\end{equation}
The saddle point is given by  the $R_i$  that solve
\begin{equation}
F^{\prime}(R_i)=-{1\over R_i^2} \left(R-E+\sum{1\over R_i}\right) \ .
\label{sadpri}
\end{equation}
Finding the solution  $F(R)$ of  (\ref{saddle}-\ref{sadpri}) is   non trivial.
If however the saddle point is symmetric,  $R_1 = R_2 = \cdots = R_K$,   the
function $F(R)$  satisfies
\begin{equation}
F(R) = \max_{R_1}  \left\{ - \ \frac{1}{2} \left( E - R - \frac{K}{R_1}
\right)^2 + K F(R_1)  \right\} \ .
\label{sadsym}
\end{equation}
Even under this assumption, we could not
find a closed expression for $F(R)$ in terms of elementary functions. One can
however
show that a solution of (\ref{sadsym}) is given by
\begin{equation}
F(R) = \lim_{n \rightarrow \infty} G_n(R)
\label{FR}
\end{equation}
where the $G_n(R)$ are defined by

\begin{equation}
G_n(R) = \max_{X_1, X_2, \cdots X_n} \left\{ - \ \frac{1}{2} \sum_{p=0}^n K^p
\left( X_p - E + \frac{K}{X_{p+1}} \right)^2 \right\}
\label{GR}
\end{equation}
with $X_0 = R$ and $X_{n+1} = A  $.

  It is  easy to show that (\ref{FR}) and (\ref{GR}) solve
(\ref{sadsym}). First,  it is clear that $G_n(R) \leq 0 $ for all $n$ and $R$.
Second, it is easy to check that $G_n(R)$ increases with $n$. This can be  seen
by choosing
the same set $X_1, X_2, \cdots X_n$ for $G_n$ and $G_{n+1}$ with  $X_{n+1}=
X_{n+2} = A  $. Since the $G_n$ increase and are all negative,
the limit (\ref{FR}) which defines $F(R)$  exists.
That  $F(R)$ satisfies (\ref{sadsym}) then follows from the observation that
\begin{equation}
G_{n+1}(R) = \max_{R_1}  \left\{ - \ \frac{1}{2} \left( E - R - \frac{K}{R_1}
\right)^2 + K G_n(R_1)  \right\}
\end{equation}

We tried to check  the assumption that the saddle point equation
(\ref{saddle}) has  a symmetric saddle point  by solving this equation
numerically for
$K=2$ and $E= 3. \ $. The values of $R_1$ and $R_2$ which give the saddle point
are shown in figure 4 and the numerical solution for
$F(R)$ is shown on figure 5.
We see that the saddle point is symmetric for a range of values of $R$  around
$A  $.
Outside this range, however, the saddle point is no longer symmetric
and $F(R)$ seems to be constant.  The rather complicated shape of $F(R)$  makes
 the calculation of the integrated density of states difficult, in particular
because it is  hard to tell if the two terms which appear in the expression
(\ref{omegar}) of $\langle \omega(E) \rangle $ are of the same order when
$\lambda $ is small, and because if the two terms are of the same order,
 the prefactor $Q(R)$ in (\ref{PQF}) would  need to be calculated. In any case,
one would find a non-zero integrated density of states $\langle \omega(E)
\rangle$ which would be exponentially small when $\lambda \rightarrow 0$ and
which would vanish to al
l orders in perturbation in $\lambda$.

\section{Numerical attempt to determine  the mobility edge}
The expression (\ref{Ec})  obtained in section 5
predicts that  the mobility edge tends to the band edge value
$E= 2\sqrt{K}$ as $\lambda \rightarrow 0$. This  disagrees with
the result of Abou-Chacra and Thouless [5]  that the mobility edge tends to $E=
K+1$ as
$\lambda$ gets small.

That the perturbative expansion   might be  insufficient to predict the
location of the mobility edge has  been discussed above: the perturbative
expansion
predicts that for $E > E_c$ the density of states is zero to all orders in
$\lambda$, whereas one
knows that the density of states never vanishes  for
an unbounded distribution  $\rho(V)$ of potentials. We tried to show in the
previous section
that a way of resolving this difficulty  is to have non-perturbative
contributions. The same could also  happen for the  reflection
amplitude: one could have
$|r|=1$ for $E>E_c$ to all orders in  $\lambda$ but
with $|r| \neq 1$ because of non-perturbative effects.

One way of locating the mobility edge is to determine whether
the distribution $P_{\rm real}(R)$ is stable against imaginary perturbations.
We know that if the $R_i$ are real at the boundary, they  remain real under
the iteration (\ref{rec}). Now let us add an infinitesimal imaginary component
$i \epsilon$
to the $R_i$ at the boundary. After a finite number $n$ of iterations of
(\ref{rec}),
the imaginary part of $R_i$ is still infinitesimal and is proportional to
$\epsilon$.
Calling this imaginary part  $i \epsilon Y_i$, we have
\begin{equation}
Y_i = \sum_{\rm paths}  \ \prod_{j \in {\rm path}} \frac{1}{R_j^2}
\label{path}
\end{equation}
where the sum in (\ref{path}) runs over all the $K^n$ paths of $n$ steps from
site
$i$ to the boundary of the tree.
If one computes the $Y_i$,  either they go to zero as $n$ increases, meaning
that
$P_{\rm real}(R)$ is stable or they grow with $n$, meaning  that $P_{\rm
real}(R)$ is unstable.

\subsection{Numerical approach}
We did not find an analytical way of determining whether the $Y_i$ grow or
decay under the iteration procedure. So we had recourse to
 a MonteCarlo method.
We represent the distribution $P_{\rm real}(R)$ by a sample of $N$ points,
where typically $N=100,1000,10000$. This means that we have $N$ values of the
$R_i$
and $N$ values of the $Y_i$.
At each elementary step, we update one $i$ chosen at
random,   by chosing $K$ indices at random
  $j_1, \cdots j_K$ between $1$ and $N$
 and a random value of the potential $V_i$ and we replace $R_i$ and  $Y_i$
by

\begin{equation}
R_i = E - \lambda V_i -  \left( \frac{1}{R_{j_1}} + \frac{1}{R_{j_2}}  \cdots +
\frac{1}{R_{j_K}} \right)
\end{equation}

\begin{equation}
Y_i =  \frac{1}{R_{j_1}^2}  Y_{j_1}  +
 \frac{1}{R_{j_2}^2} Y_{j_2}
 \cdots + \frac{1}{R_{j_K}^2} Y_{j_K} \ .
\label{YY}
\end{equation}
We start with the $Y_i$ of order $1$ and we iterate this procedure until all
the $Y_i$ have become either very large $10^{30}$ or very small $10^{-30}$.
The mobility edges estimated by this procedure are shown on figure 6 and 7
for the Gaussian and Cauchy distributions respectively. The points for
different values of $\lambda$ are obtained for different samples, and so the
roughness of the curves indicates the statistical errors.

We  expect this procedure to give the true mobility edge for $N$ infinitely
large since for large $N$ the values of $R_{j_1}, R_{j_2}, \cdots R_{j_K}$ are
independent.
We see that as $N$ increases the estimated mobility edge moves upward and to
the right, and so it is not so easy to predict from these data accurate values
of the large $N$ limit.
For a Gaussian potential and for small $\lambda$, the results
shown in figure 7b seem to agree well with the expression (\ref{Ec}).
Notice however that the agreement becomes  worse as $N$ increases.
This means that even though
the mobility edge in the limit $\lambda \rightarrow 0$ starts
at $2 \sqrt{K}$ and not $K+1$ in our simulations,
we cannot conclude from the data that the mobility edge really
starts at $2 \sqrt{K}$  when $\lambda \rightarrow 0$ because the limits $N
\rightarrow \infty$ and $\lambda \rightarrow 0$ may not commute.

\subsection{The case of independent $R$}
In order to test the validity of this numerical approach, it is useful to try
it in an exactly soluble case.
If  we suppose the $R_i$ to be independent random variables
distributed according to a given probability distribution
$P_{\rm real}(R)$, one knows from the theory of directed polymers [32,33]  on a
tree
the exact expression of the large $n$ limit of $\log Y_i /n$ where the $Y_i$
are defined by (\ref{path}). To test the MonteCarlo procedure descibed above,
we computed
the "mobility edge"  by  using   (\ref{YY})
 with independent  $R_i$  chosen according to  their exact probability
distribution (\ref{PCauchy}). Then using the known results from the problem of
directed polymers one has for large $n$,
\begin{equation}
(Y_i)_{\rm typical} \simeq  \left[ \  \min_\beta \left( K \langle R^{-2 \beta}
\rangle
  \right)^{\frac{1}{\beta}}  \ \right]^n
\end{equation}
The line in the plane $E, \lambda$  which separates the region of very large
$Y_i$ and very small $Y_i$ can then be obtained exactly using the following
expression for
$\langle R^{-2 \beta} \rangle $.
\begin{equation}
\langle R^{-2 \beta} \rangle = \frac{b}{\pi} \int_{-\infty}^\infty dR
\frac{R^{-2 \beta}}{(R-a)^2 + b^2} = \frac{b^{-2 \beta}}{ \cos(\pi \beta)}
\left( \frac{b^2}{a^2+b^2}\right)^\beta \cos \left( 2 \beta \tan^{-1} \left(
\frac{a}{b} \right) \right)
\end{equation}
The exact curve for  independent $R_i$ is shown in figures 6c and 7c
together with the results
of the MonteCarlo procedure. We see that  the results seem to converge rather
slowly as $N$ increases. Moreover, the limits $N \rightarrow \infty$ and
$\lambda \rightarrow 0$ do not
seem to commute, as the exact result tends to $K+1$ as predicted by Abou-Chacra
and Thouless [5]
whereas the finite $N$ results tend to the band edge value $2 \sqrt{K}$.

In summary we see that the MonteCarlo procedure described above can in
principle be used to determine the mobility edge. However, $N=10000$ does not
seem to be  large enough  to resolve numerically the question of where the
mobility edge starts in the $\l
ambda \rightarrow 0 $ limit.

\section{Conclusion}
We have shown that
a great deal of information on the Anderson model on  a tree is
is contained in the random recursion (\ref{rec}).
If we assume that
under iteration of (\ref{rec}), the distribution of $R$ converges to a fixed
distribution $P(R)$,  the problem of knowing whether the wave functions are
extended or localised reduces to the question of the existence of a
complex fixed distribution $P_{\rm complex}(R)$. The equation
  (\ref{Pf}) that
$P(R)$ satisfies is in fact equivalent to
those already obtained by other methods [5-9].
We think
however that our way of deriving these equations is  more direct.

The main result of this work was
to show that one can expand quantites of interest
like the density of states, the mobility edge or the reflection amplitude in
powers of $\lambda$. This approach is not however entirely satisfactory
because we have not fully understood the non-perturbative effects.
In particular, the existence and nature
of non-perturbative corrections to the
complex fixed distribution $P_{\rm complex}(R)$
remains an open question, as, by consequence, does the position
of the mobility edge in the limit $\lambda\to 0$.
Both our perturbative expansion and our MonteCarlo simulations
indicate that the mobility edge starts at the band edge
of the pure system, but non-perturbative effects could change the
former prediction, and as is the case for the independent $R$ of section 7,
the limit $N\to \infty$ and the limit $\lambda\to 0$ may not commute in the
latter.

We think several important points deserve further consideration.

First,  it would be nice to be able to mathematically
prove that the sequence
of $P_n(R)$ converges and to know under what conditions
a complex fixed distribution exists.
This problem is not  easy  because, as we discussed above,
in the pure case,
the sequence does not converge for energies inside the band and somehow it is
the effect of a weak disorder which makes the distribution concentrate aroud
the complex fixed point.

Second, it would be interesting to develop a
non-perturbative approach, especially for the complex distribution,
in order to compute at
least for small $\lambda$ the shape of the mobility edge.
Despite our efforts, we were unable to find a method
allowing us to describe $P_{\rm complex}(R)$ for small $\lambda$ in the range
of energies $|E| > 2 \sqrt{K}$.

One could also try to use the recursion (\ref{rec})  to calculate
other quantities which play an important role in the localisation problem, such
as the inverse participation ratio.

Last, we think that the
MonteCarlo procedure described in section 7 could be used to
accurately determine the position of the mobility edge
by increasing $N$ to 1 or 10 million, though this would
require a rather serious numerical effort.
\\ \\ \\

\appendix{}
\setcounter{equation}{0}
\def\theequation{A\arabic{equation}}

In this appendix we
derive the relation (\ref{pnrc}) between the real and complex
fixed point distributions and the expression (\ref{omegac}) of the integrated
density
of states from the expression (\ref{omegar}).
To do so let us asume that we have a sequence of
complex distributions $P_{\rm complex}^n(r-is)$ concentrated on the lower half
plane
which satisfy the recursion (\ref{P2})).
To each of these distributions we associate a real
distribution $Q^n(R)$ by
\begin{equation}
Q^{n}(R)= \frac{1}{\pi} \int_{-\infty}^{\infty} dr \int_{0}^{\infty}d s
{s\over (R-r)^2+s^2} \  P^{n}_{\rm complex}(r-is).
\label{Qn}
\end{equation}
We are going to show that the sequence of $Q^n(R)$ thus defined
also satisfies (\ref{P2}).

 From the recursion relation (\ref{P2}) for the complex distribution $ P_{\rm
complex}^n$,
we can rewrite (\ref{Qn}) as
\begin{eqnarray}
 Q^{n+1}(R)= &&\int_{-\infty}^{\infty} dr \int_{0}^{\infty} ds \
\prod_{i=1}^k\int_{-\infty}^{\infty}dr_i\int_{0}^{\infty}ds_i
{1\over \pi}{s\over (R-r)^2+s^2} \int  \rho(V) dV
 \label{QQ}
 \\ && \ \ \ \ \  P^n_{\rm complex}(r_i-is_i)\
\delta (r-E+\lambda V+\sum_{i=1}^k{r_i\over r_i^2+s_i^2})
 \ \delta (s-\sum_{i=1}^k{s_i\over s_i^2+r_i^2})
\nonumber
\end{eqnarray}
It is useful now to note two
properties of Cauchy distributed random variables.
A Cauchy distribution is a probability
distribution $C(x;a,b)$ defined by
\begin{equation}
C(x;a,b)={1\over \pi}{b\over (x-a)^2+b^2}
\end{equation}
where  $a$ is real  and $b$ is real and positive.
Let $x_1$ and $x_2$ be Cauchy distributed
random variables with distributions
$C(x_1;a_1,b_1)$ and $C(x_2;a_2,b_2)$  respectively.
Then the sum $x=x_1+x_2$ is
Cauchy with distribution
$C(x;a_1+a_2,b_1+b_2)$. Similarly, if $x$ is distributed according to
$C(x;a,b)$, the inverse $y={1\over x}$
is Cauchy with distribution
$C(y; a/(a^2+b^2), b/(a^2+b^2) )$.
Hence, if the real variables $R_1,...,R_k$ are
Cauchy distributed with distributions
$C(R_i; r_i,s_i)$, it follows   for fixed $V$ that the
real variable $R$ defined by
\begin{equation}
R=E-\lambda V-\sum_{i=1}^k{1\over R_i}
\end{equation}
is distributed according to $C(R;r,s)$
where $r=E-\lambda V-\sum_{i=1}^k r_i/( r_i^2+s_i^2)$
and $s=\sum_{i=1}^k s_i/( s_i^2+r_i^2)$.
Written out as an equation, this reads
\begin{equation}
{1\over \pi}{s\over (R-r)^2+s^2}=
\int_{-\infty}^{\infty} \ \prod_{i=1}^k dR_i \
{1\over \pi} \ {s_i\over (R_i-r_i)^2+s_i^2} \ \delta (R-E+\lambda
V-\sum_{i=1}^k
{1\over R_i})\label{70}
\end{equation}
If we substitute (\ref{70})
 into (\ref{QQ}),
and use (\ref{Qn}) namely that
\begin{equation}
Q^n(R) =\int_{-\infty}^{\infty}dr_i\int_{0}^{\infty}ds_i
{1\over \pi}{s_i\over (R-r_i)^2+s_i^2}P^n_{\rm complex}(r_i-is_i)
\end{equation}
it follows that the $Q_n(R)$ also obey (\ref{P2}).
Assuming that in the limit $n\rightarrow \infty$
the $Q^n(R)$ converge establishes (\ref{pnrc}).
\\ \\ \\
Let us now see how (\ref{omegac}) can be obtained from (\ref{omegar}).
The expression (\ref{omegar}) can be rewritten as
\begin{eqnarray}
\langle \omega(E) \rangle =
  - \frac{1}{\pi} {\rm Im} \left\{   \frac{ K-1}{2}\int_{-\infty}^{\infty} \
P_{\rm real}(R) \  dR  \int_{-\infty}^{\infty} \   P_{\rm real}(R') \   dR' \
\log( R'-i \epsilon - \frac{1}{R-i \epsilon} )   \right.
\nonumber \\
\left. + \frac{K+1}{2} \int_{-\infty}^\infty P_{\rm real}(R) \  dR  \ \log(R -
i \epsilon) \right\}
\end{eqnarray}
where $- i \epsilon$ is an infinitesimal imaginary part.
One can then replace $P_{\rm real}(R)$ in this expression by
 (\ref{pnrc}). Using the residue theorem to
do the integral over the real
variable $R$ gives (\ref{omegac}).

\ack
We thank  M. Aizenman and S. Ruffo for useful discussions.
\\ \\
\\ \\
\newpage
\figures
\begin{description}
\item
Figure 1: A Cayley tree with $K=2$ and depth n=4.
\item
Figure 2: A branch of a Cayley tree with wires attached to the
boundaries to test the conducting properties. In the scattering situation, a
plane wave is sent
in from the left and allowed to scatter off the tree.
\item
Figure 3: A set of typical $R$
obtained by
 the Monte-Carlo version of the
recursion (\ref{rec}) explained in section 7 for $N = 1000$.
Here $K=4$, $E=2.1$, and the complex fixed point solution of (\ref{Astar}),
represented by a
diamond, lies at $A=1.05-1.702 i$.
The ellipse represents the set of points to which the
initial $R=2-i$ would map in the absence of disorder.
For small disorder, $\lambda=.1$ (fig 3a), the $R$ quickly
move off the ellipse and concentrate themselves around the
complex fixed point. As the disorder increases, $\lambda=1.$  (fig 3b)
and $\lambda=5.$ (fig 3c), the $R$ spread out. Finally,
for very large disorder,
$\lambda=30.$ (fig 3d), the $R$ quickly become real,
and the particle is localised.
\item
Figure 4:   The solutions $R_1(R)$
and $R_2(R)$ of the saddle point equation (\ref{saddle})
for $K=2$ and $E=3.$ as a function of $R$.
Near the stable fixed point $A$, the saddle point is symmetric:
$R_1(R)=R_2(R)$, while for larger and smaller values of $R$ the
symmetry is broken, $R_1(R)\ne R_2(R)$.
\item
Figure 5: Numerical solution of the saddle point equation (\ref{saddle})
for $F(R)$ with $K=2$ and $E=3. $ \ . The function $F(R)$ seems to be constant
in the range where the symmetry between $R_1(R)$ and $R_2(R)$ is broken.

\item
Figure 6a: Monte Carlo
determination as described in section 7 of the mobility edge for
Cauchy distributed disorder. Here $K=4$.
The number $N$ of different $R$ is $100, 1000, 10000$.
Figure 6b: Same as figure 6a for the Gaussian disorder.
Figure 6c: Same as figure 6a for \underline{ independent}
$R$ distributed according to the Cauchy distribution (\ref{PCauchy})
\item
Figures 7a,7b,7c: Enlargements of figures 6a, 6b and 6c
showing small values of $\lambda$.
Figure 7b shows a rather nice agreement with the result (\ref{Ec}).
However as $N$ increases the agreement seems to become worse indicating
 that the limits $N \rightarrow \infty$
and $\lambda \rightarrow$ do not commute. This is even more apparent
for the independently distributed $R$ where
the exact result converges in the limit $\lambda \rightarrow 0$
to $E= K+1$ as predicted by Abou-Chacra and Thouless [5]
whereas the finite $N$ curves converge to the band edge $2 \sqrt{K}$.

\end{description}

\references
\refjl{[1] Anderson P W }
{ Phys. Rev.}{109}{1492 (1958)}
\refjl{[2] Thouless D J}
{Phys. Rep.}{13}{93 (1973)}
\refbk{ [3] Souillard B  in Chance and Matter }
{Les Houches XLVI, 1986}{ed by J Souletie, J Vannimenus and R Stora (1987)}
\refjl{[4] Bulka B, Kramer B and MacKinnon A}
{Z. Phys.B}{60}{13 (1985)}
\refjl{ [5] Abou-Chacra R, Anderson P W and Thouless D J}
{J. Phys. C}{6}{1734 (1973)}
\refjl{[6] Abou-Chacra R and Thouless D J}
{J. Phys. C}{7}{65 (1974)}
\refjl{ [7] Kunz H and Souillard B}
{J.  de Physique(Paris) Lett.}{44}{L411 (1983)}
\refjl{ [8] Mirlin A D and Fyodorov Y V}
{Nucl. Phys. B}{366}{507 (1991)}
\refjl{ [9] Mirlin A D and Fyodorov Y V}
{J. Phys. A}{24}{2273 (1991)}
\refjl{[10] Kim Y and Harris A B}
{Phys. Rev. B}{31}{7393 (1985)}
\refjl{ [11] Acosta V and Klein A}
{J. Stat. Phys.}{69}{277 (1992)}
\refjl{ [12]  Kawarabayashi T and Suzuki M}
{J. Phys. A}{26}{5729 (1993)}
\refjl{[13] Efetov K B}
{Sov. Phys. JETP }{61}{606 (1985)}
\refjl{ [14] Shapiro B}
{Phys. Rev. Lett.}{50}{747 (1983)}
\refjl{[15] Chalker J T and Siak S}
{J. Phys. Cond. Matt.}{2}{2671 (1990)}
\refjl{[16] Derrida B and Rodgers G J}
{J. Phys. A}{26}{L457 (1990)}
\refjl{[17] Derrida B and Gardner E}
{J.  de Physique(Paris) }{45}{1283 (1984)}
\refjl{ [18] Wegner F}
{Z. Phys.B}{44}{9 (1981)}
\refbk{ [19] Spencer T in Critical Phenomena, Random systems , Gauge Theories}
{Les Houches XLIII, 1984}{ed by K Osterwalder and R Stora (1986)}
\refbk{ [20] Pastur L and Figotin A  : Spectral properties of disordered
systems
in the one-body approximation}
{Springer Verlag (1991)}
\refjl{[21] Rodgers G J and Bray A J}
{Phys. Rev. B}{37}{3557 (1988)}
\refjl{[22] Rodgers G J and De Dominicis C}
{J. Phys. A}{23}{1567 (1990)}
\refjl{ [23] Fyodorov Y V, Mirlin A D and Sommers H J}
{J. de Physique I (France) }{2}{1571 (1992)}
\refjl{ [24] Dhar D  and Ramaswamy R}
{Phys. Rev. Lett.}{54}{1346 (1985)}
\refjl{ [25] Aizenman M and Molchanov S}
{Comm. Math. Phys.}{157}{245 (1993)}
\refjl{ [26] Aizenman M}
{preprint 1993}{}{}
\refbk{ [27] Luck J M: Syst\`emes d\'esordonn\'es unidimensionnels}
{Al\'ea, Saclay}{(1992)}
\refjl{[28] Kappus M and Wegner F}
{Z. Phys.B}{45}{15 (1981)}
\refjl{[29] Lambert C J}
{ Phys. Rev. B}{29}{1091 (1984)}
\refjl{[30] Bovier A and Klein A}
{ J Stat Phys}{51}{501 (1988)}
\refjl{[31] Campanino M and Klein A}
{Comm. Math. Phys.}{130}{441 (1990)}
\refjl{[32] Buffet E, Patrick A and Pule J V  }
{ J. Phys. A}{26}{1823 (1993)}
\refjl{ [33] Derrida B, Evans M R and Speer E R}
{Comm. Math. Phys.}{156}{221 (1993)}
\end{document}